\documentstyle[preprint,aps,psfig]{revtex}
\begin{document}

\tighten
\title{Dependence of direct neutron capture on nuclear-structure models}

\author{T. Rauscher}
\address{Institut f\"ur Physik, Universit\"at Basel, Klingelbergstr.82,
CH-4056 Basel, Switzerland}

\author{R. Bieber, H. Oberhummer}
\address{Institut f\"ur Kernphysik, Technische Universit\"at Wien,
Wiedner Hauptstr. 8-10, A-1040 Wien (Vienna), Austria}

\author{K.-L. Kratz}
\address{Institut f\"ur Kernchemie, Universit\"at Mainz, Fritz-Strassmann-Weg 2,
D-55099 Mainz, Germany}

\author{J. Dobaczewski}
\address{Institute of Theoretical Physics, Warsaw University,
   Ho\.za 69,  PL-00681 Warsaw, Poland}

\author{P. M\"oller\thanks{Permanent address: P. Moller Scientific
Computing and Graphics, Inc., P.O. Box 1440, Los Alamos, NM 87544}}
\address{Institut f\"ur Kernchemie, Universit\"at Mainz,
Fritz-Strassmann-Weg 2,
D-55099 Mainz, Germany}

\author{M.M. Sharma}
\address{Physics Department, Kuwait University, Kuwait 13060}

\maketitle

\begin{abstract}
The prediction of cross sections for nuclei far off stability is crucial
in the field of nuclear astrophysics.
We calculate direct neutron capture on the even-even isotopes $^{124-145}$Sn
and $^{208-238}$Pb with energy levels,
masses, and nuclear density distributions
taken from different nuclear-structure models.
The utilized structure models are a Hartree-Fock-Bogoliubov model,
a relativistic mean field theory, and a macroscopic-microscopic model
based on the finite-range droplet model and a folded-Yukawa
single-particle potential.
Due to the differences in the resulting neutron separation and level
energies, the investigated models yield capture cross sections sometimes
differing by orders of magnitude. This may also lead to differences in the
predicted astrophysical r-process paths.
\end{abstract}


\section{Introduction}
\noindent
Explosive nuclear burning in astrophysical environments produces
unstable nuclei which again can be targets for subsequent reactions.
Most of these nuclei are not accessible in terrestrial laboratories or not fully
explored by experiments, yet.

Approximately half of all stable nuclei observed in nature in the heavy
element region $A>60$ were produced in the so-called r-process (i.e.,
rapid neutron capture process), which is believed to occur
in type II supernova explosions (see e.g., \cite{cowan,ohurev}).
An environment with a high neutron density is the prerequisite for such an
r-process, in which heavier elements are built up from seed elements by
consecutive neutron captures and $\beta$-decays. Because of the abundant
neutrons, a multitude of neutron captures ($\simeq 15-35$)
may occur until the $\beta$-decay
half-life becomes shorter than the half-life against neutron capture. Thus,
the r-process path along which reactions take place, is pushed off the
region of stability towards neutron-rich unstable nuclei. The location of
the path has consequences for the resulting nuclear abundances, calculated in
astrophysical models~\cite{friedel,chen96}.

For most of the required neutron capture cross sections the statistical
model (compound nucleus (CN) mechanism, Hauser-Feshbach approach)
can be applied. This
model employs a statistical average over resonances, for which one has to
know level densities but not necessarily exact excitation energies and level
spin assignments. However, the criterion for the applicability of that model
is a sufficiently high level density. Especially for some light nuclei it
has been known for years that the statistical model cannot be applied and that
the direct capture (DC) mechanism dominates the cross sections. Nevertheless,
it has only been realized recently that also for intermediate and heavy 
nuclei the
direct mechanism can become important near shell closures and for neutron-rich
isotopes when the level density becomes too low for the CN
mechanism. When approaching the drip-line, neutron separation energies
decrease and the nuclei become less deformed, both leading to a smaller
level density at the relevant projectile energy. This relevant energy is
determined by the peak
$E=kT$ of the Maxwell-Boltzmann velocity distribution of the neutron gas.
If a segment of the r-process path at a given element lies close enough
to the drip-line, the statistical model will not be applicable anymore and
the DC reactions will dominate~\cite{mat83,gor97}.

The relation between DC and CN mechanisms has already been studied
for neutron capture by light
and intermediate target nuclei~\cite{ohurev,bal,gru,mei,meis,bee,kra}.
Investigations of the dependence of the level density on charge and mass number
and a discussion of the applicability of the statistical model have been
given elsewhere~\cite{tfkl1}. In this paper we want to investigate
direct neutron capture on neutron-rich Sn and Pb isotopes with the emphasis
on discussing the difficulties, the level of reliability as well as
the predictive power of
theoretical calculations.

The main problem for the DC predictions is that neutron separation
energies and level properties (excitation energies, spins, parities)
have to be
known accurately, contrary to a statistical calculation in which it is
sufficient to know the level density. As
in the foreseeable future one can not expect
any experimental
information for the majority of nuclei close to the drip-line,
one has to turn to theory for providing the input for the DC
calculations. At the moment, there are several microscopic and
macroscopic-microscopic descriptions competing in the quest for predicting
nuclear properties far off stability. For the first time, in this work we
want to investigate the difference in the level structure between several
models and its impact on predicted neutron capture cross sections. The
compared models are a Hartree-Fock-Bogoliubov (HFB) model with the Skyrme SkP
force~\cite{doba1,doba}, a relativistic mean field theory (RMFT) with the 
parameter set NLSH~\cite{sharma1,sharma2}, and the macroscopic-microscopic
finite-range droplet model FRDM (1992) which was also used in
calculations of nuclear ground-state masses and 
deformations~\cite{moeller,moell2} and in calculations of quantities of
astrophysical interest~\cite{moellkra}.

In Section~\ref{secDC} we very briefly introduce the method of the DC
calculation and Section~\ref{secMic} gives an overview of the utilized
microscopic models. For $^{208}$Pb, the DC results can directly be compared to
experimental values. This is described in Section~\ref{secExp}.
In the following Sections~\ref{secPb} and ~\ref{secSn} we present our results
for the heavy Pb and Sn isotopes. Possible astrophysical signatures and
remaining uncertainties are discussed in Section~\ref{SecDis}.
The paper is concluded by the summary section~\ref{summary}.

\section{Direct Capture and Folding Procedure}
\label{secDC}
\noindent
The theoretical cross section $\sigma^{\mathrm{th}}$ is derived from the
DC cross section $\sigma^{\mathrm{DC}}$ given by~\cite{kra,kim}
\begin{equation}
\sigma^{\mathrm{th}}=\sum_{i}C_i^2S_i\sigma^{\mathrm{DC}}_i \quad .
\end{equation}
The sum extends over all possible final states (ground state and
excited states) in the residual nucleus. The isospin Clebsch-Gordan
coefficients and spectroscopic factors are denoted by $C_i$ and $S_i$,
respectively. The DC cross sections $\sigma^{\mathrm{DC}}_i$ are
essentially determined by the overlap of the scattering wave function in
the entrance channel, the bound-state wave function in the exit channel,
and the multipole transition operator. For the computation of the DC
cross section we used the direct capture code TEDCA~\cite{TEDCA}, which
includes E1, M1 and E2 transitions.

For determining the nucleon-nucleus potential the folding procedure was
employed, a method already successfully applied in the description of
many systems. In this approach the nuclear target
density $\rho_{\mathrm{T}}$ is folded
with an energy and density dependent nucleon-nucleon interaction
$v_{\mathrm{eff}}$~\cite{kob}:
\begin{equation}
V(R)=\lambda V_{\mathrm{F}}(R)=\lambda \int \rho_{\mathrm{T}}
(\bbox{r})v_{\mathrm{eff}}(E,\rho_{\mathrm{T}},
\vert\bbox{R} - \bbox{r} \vert )d\bbox{r} \quad ,
\end{equation}
with $\bbox{R}$ being the separation of the centers of mass of the two
colliding nuclei. The normalization factor $\lambda$ accounts for
effects of antisymmetrization and is close to unity. The nuclear density
$\rho_{\mathrm{T}}$ can be derived from experimental charge
distributions or from theory. The potential
obtained in this way ensures the correct behavior of the wave functions
in the nuclear exterior. At the low energies considered in astrophysical
events the imaginary parts of the optical potentials are small.

In connection with the results presented below it is useful to recapitulate
the sensitivity of the DC calculations to various elements of the
description. In ascending importance,
in the present context the DC is sensitive to the optical potential and
density distribution, respectively, the reaction $Q$-value, and the spin and
parity of a level.

For the accuracy attempted here, there is almost no
difference in the results obtained by employing the optical potentials
derived from the density distributions of the different models while leaving
all other properties unchanged.

A stronger dependence is seen when examining changes in the $Q$-value. An 
increase
in the $Q$-value will give a non-linear increase in the resulting cross section.
As the $Q$-value is computed as the difference in the binding energies of
target and residual nucleus (i.e., the neutron separation energy) minus the
excitation energy of the level into which the neutron is captured
\begin{equation}
Q_i=(B_{\mathrm{T}}-B_{\mathrm{R}})-E_i=S_{\mathrm{n}}-E_i\quad,
\end{equation}
the cross section will be sensitive to the masses (separation energies) derived
in the different microscopic models as well as the level structure (excitation
energies) given in these models.

The by far strongest sensitivity is that to spins and parities of the
involved initial and final states. In
order to comply with the electromagnetic selection rules, a state has to
have the proper parity to contribute to the cross section significantly.
The dominant contribution to the DC cross section will stem from an E1
transition. In this case, parity has to change.
Consequently, the capture of an incoming neutron $p$-wave will be important 
for the Pb isotopes, whereas
$s$-wave capture is dominating in the Sn cases.
Furthermore, significant contributions only arise from low spin states
like 1/2 and 3/2 states, whereas the capture to levels with higher spins is 
strongly suppressed.
In this respect, it will prove to be important that the different
microscopic models make different predictions on which states are
neutron-bound and which are not, since DC can only populate bound states.

\section{The Microscopic Input}
\label{secMic}
\noindent
The energy levels, masses, and nuclear density distributions
needed as input for the DC calculation were
taken from three different approaches. The first one was the
RMFT which has turned out to be a successful tool
for the description of many nuclear properties~\cite{gam}. The RMFT
describes the nucleus as a system of Dirac nucleons interacting via
various meson fields. There are six parameters which are usually
obtained by fits to finite nuclear properties. For our calculations we
have used the parameter set NLSH~\cite{sharma1,sharma2}.

The second method was FRDM (1992), which is a macroscopic-microscopic
model based on the finite-range droplet macroscopic model and a
folded-Yukawa single-particle potential~\cite{moeller}.
For pairing, the
Lipkin-Nogami pairing model~\cite{yuk} is employed. This model proved to
be very successful in reproducing ground state spins along magic
numbers~\cite{moell1} and
has been used in QRPA calculations of $\beta$-decay half
lives~\cite{moell1,moellkra} and
for nuclear mass determinations~\cite{moell2}.

Finally, we also utilized the self-consistent mean field HFB
model~\cite{doba1,doba} in which the nuclear states are calculated by a one-step
variational procedure minimizing the total energy with respect
to the occupation factors and the single-particle wave functions simultaneously.

To be able to compare the predictions from all of the models the nuclei were
considered to be spherically symmetric. The limitations of such a
restriction
are discussed in Section \ref{SecDis}.

\newcommand{\Pbngamma}{$^{208}$Pb(n,$\gamma$)$^{209}$Pb}
\section{Comparison with Experiments for the {\protect\Pbngamma} reaction}
\label{secExp}

\noindent
Recently, it became possible to extract the non-resonant part of the
experimental capture cross section for the
$^{208}$Pb(n,$\gamma$)$^{209}$Pb reaction~\cite{corvi}. In that work, high
resolution neutron capture measurements were carried out in order to
determine twelve resonances in the range 1--400 keV. From these values the
resonant Maxwellian-averaged cross section $<\sigma>^{\mathrm{R}}_{30
\mathrm{keV}}$=0.221(27)\,mb was calculated.
Measurements of the total cross section using neutron
activation~\cite{macklin,ratzel} are also available at 30 keV,
yielding the value
$<\sigma>^{\mathrm{t}}_{30\mathrm{keV}}$=0.36(3)\,mb. By a simple
subtraction of the
resonant part from the total cross section the value of
$<\sigma>^{\mathrm{NR}}_{30\mathrm{keV}}$=0.14(4)\,mb can be deduced
for the
non-resonant capture cross section.

Using the experimentally known density distributions~\cite{devries},
masses~\cite{audi} and energy levels~\cite{nucldatasheets}, we
calculated the non-resonant contribution in the DC model. The strength
parameter $\lambda$ of the folding potential in the neutron channel was
fitted to experimental scattering data at low energies~\cite{scdata}.
The value of $\lambda$ for the bound state is fixed by the requirement
of correct reproduction of the binding energies.
The spectroscopic factors for the relevant low lying states of $^{209}$Pb
are close to unity as can be inferred from
different $^{208}$Pb(d,p)$^{209}$Pb reaction data~\cite{nucldatasheets}.
For the Maxwellian-averaged non--resonant DC cross section we obtained
$<\sigma>^{\mathrm{DC}}_{30\mathrm{keV}}=0.135$ mb, which is in excellent
agreement with experiment. The by far highest contributions to the DC
cross section come from the E1 $p$-wave capture to the low spin states
$J^{\pi}=1/2^+, 3/2^+, 5/2^+$. Capture to the other states is negligible.

\begin{figure}
\psfig{file=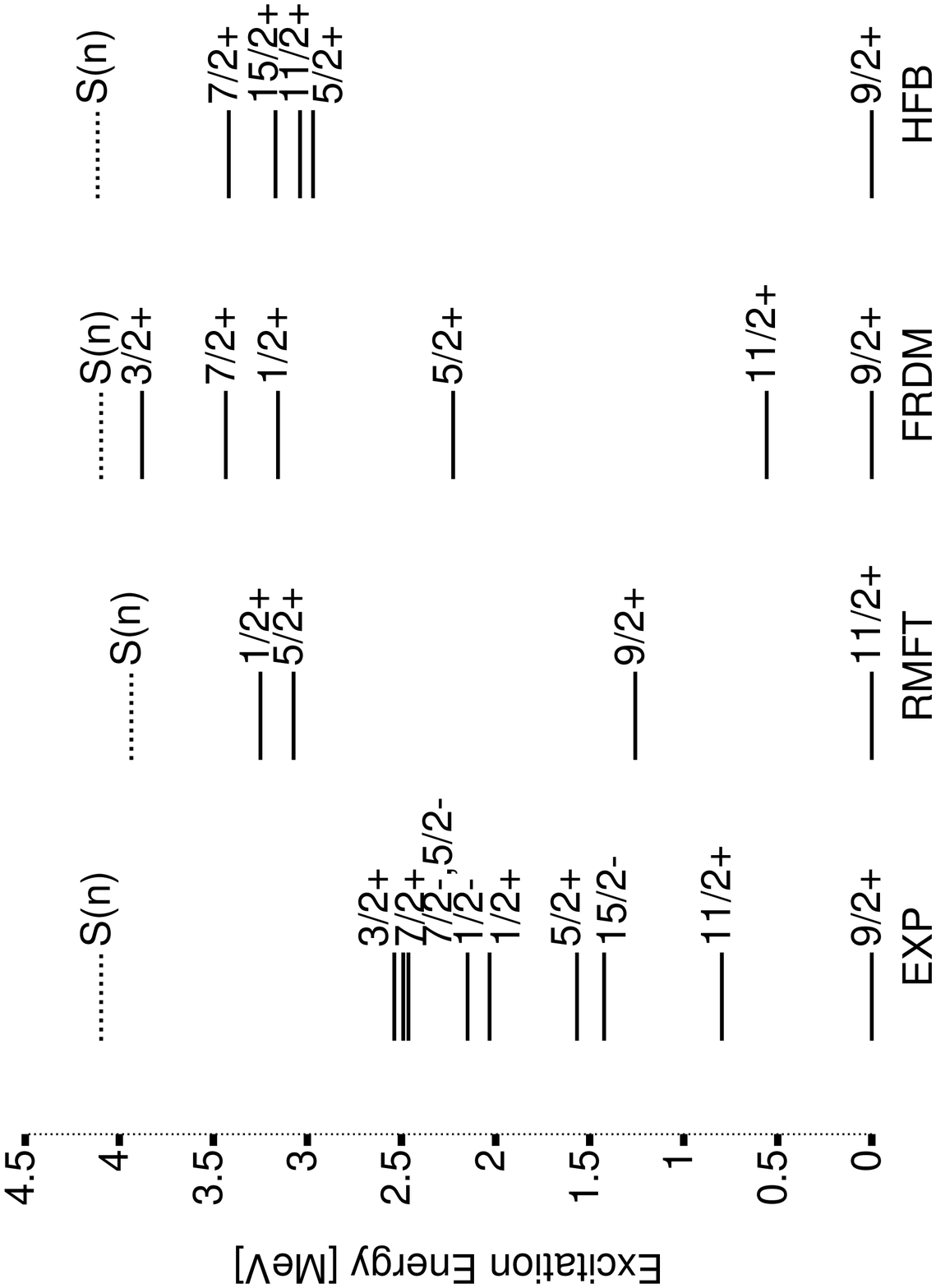,width=15cm}
\caption{\label{209}Level schemes of $^{209}$Pb obtained from experiment (EXP),
and within the RMFT \protect\cite{sharma2},
FRDM \protect\cite{moeller}, and HFB
\protect\cite{doba1,doba}.}
\end{figure}

In order to test the different microscopic approaches we also calculated
non-resonant DC on $^{208}$Pb by consistently
taking the input (energy levels, masses
and nuclear
densities) from the models described above.
Again, the strength parameter $\lambda$ of the folding potential in the
entrance channel was adjusted to the elastic scattering data for each
of the models. The calculations for the neutron capture cross
sections yield  0.0289 mb, 0.0508 mb, and 0.0135 mb
for RMFT, FRDM, and HFB, respectively.
Hence, each of the models gives a smaller value for
the Maxwellian-averaged 30 keV capture cross section than the
calculation using experimental input data. The differences are due to the
neutron separation energies and level schemes of the relevant states in
$^{209}$Pb (see
Fig.~\ref{209}) in the microscopic models, leading to different $Q$-values
for capture to the
excited states ($J^{\pi}=1/2^+,3/2^+,5/2^+$).
It should be noted that in Fig.~\ref{209} only those theoretical levels
are shown which contribute to the cross section, i.e. only particle states.
Capture into hole states is strongly suppressed by the fact that a re-ordering
process would be required in the final nucleus (see e.g.~\cite{tomenam} for
a similar case). This would be reflected in
extremely small spectroscopic
factors. Therefore, the DC to such states is negligible.

\section{Results for Neutron-rich Pb Isotopes}
\label{secPb}
\noindent
We also investigated the model dependence of neutron capture on the
neutron-rich even-even isotopes $^{210-238}$Pb. For these isotopes
experimental data are only available near the region of stability. For
more neutron-rich nuclei one has to rely solely on input parameters from
microscopic models. In this and the following section we compare cross
sections calculated with the nuclear properties predicted by different
nuclear-structure models. Therefore, we consider nuclear cross sections
instead of Maxwellian-averaged ones as in the previous section.

Having obtained the relevant spins and calculated the $Q$-values from
the masses as
discussed above, we still had to determine the scattering potentials
with their respective strength parameters
(see Eq.~2). As a first
step, the folding potentials were calculated, using the density
distributions taken from the three different nuclear-structure models
(HFB, RMFT, FRDM).
In the potentials for each of the isotopes a factor $\lambda$ was chosen
giving the
same volume integral as for the fitted $^{208}$Pb+n potential, which was
obtained as described in the previous section. This is justified because
it is known that the
volume integrals only change very slowly when adding neutrons to a
nucleus~\cite{satchler}.
For the bound
state potentials $\lambda$ is fixed by the requirement of correct
reproduction of the binding energies. The spectroscopic factors were
assumed to be unity for all transitions considered.

\begin{figure}
\psfig{file=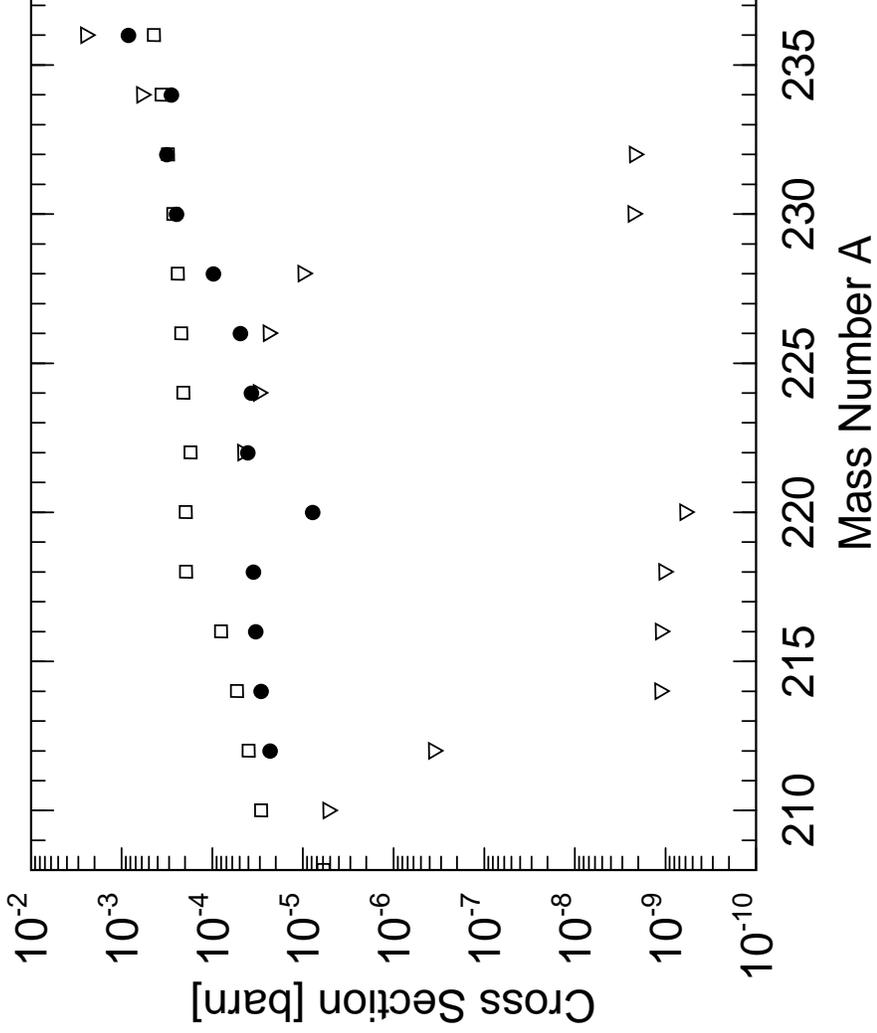,width=15cm,rheight=21cm}
\caption{\label{vgl}Direct-capture cross sections at 30
keV for different Pb isotopes.
Levels and
masses are calculated within the RMFT
(triangles),
FRDM (dots), and HFB
(squares).}
\end{figure}

\begin{figure}
\psfig{file=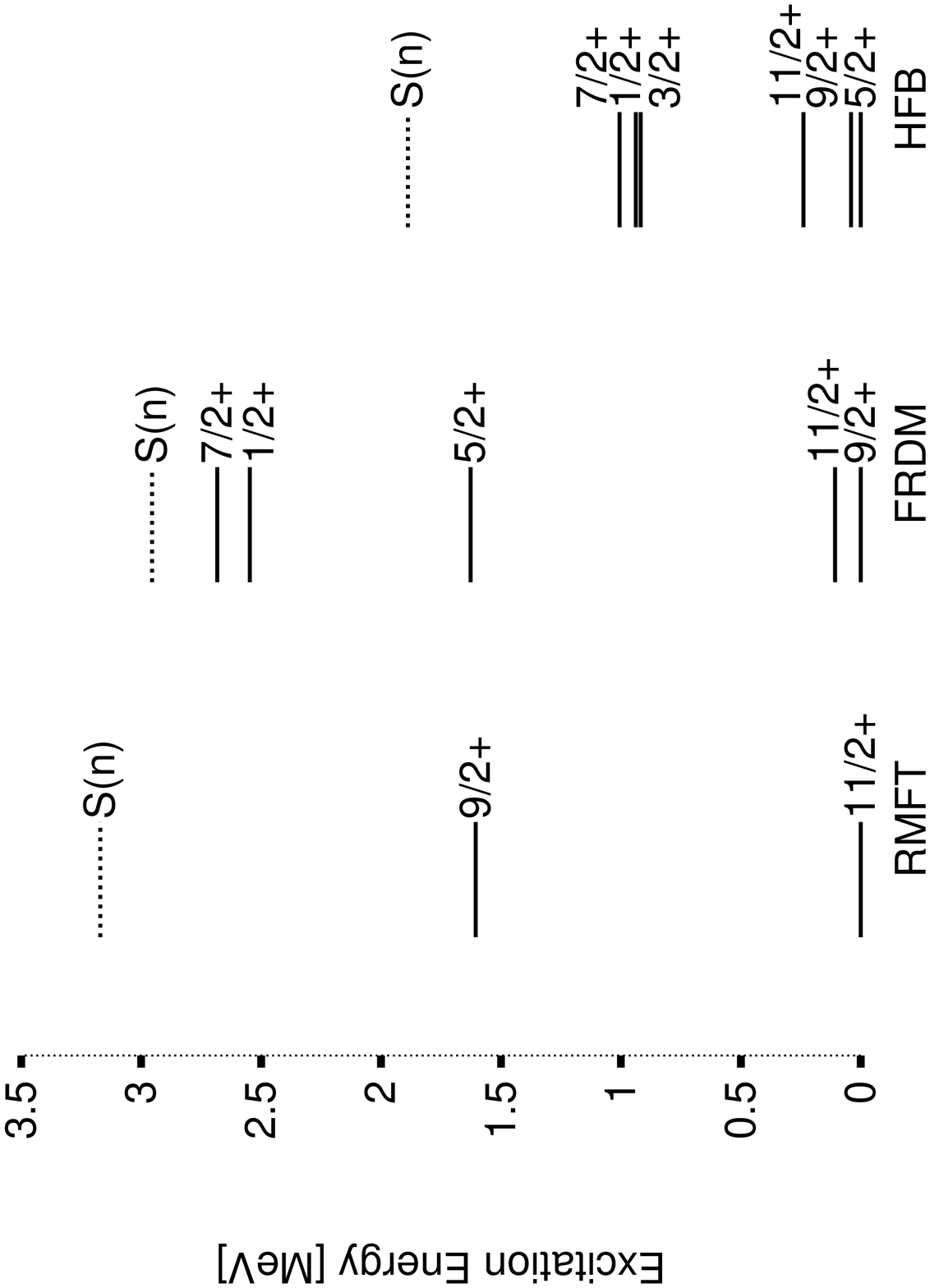,width=15cm}
\caption{\label{level1}Level schemes of $^{219}$Pb calculated
 within the RMFT,
FRDM, and HFB.}
\end{figure}

\begin{figure}
\psfig{file=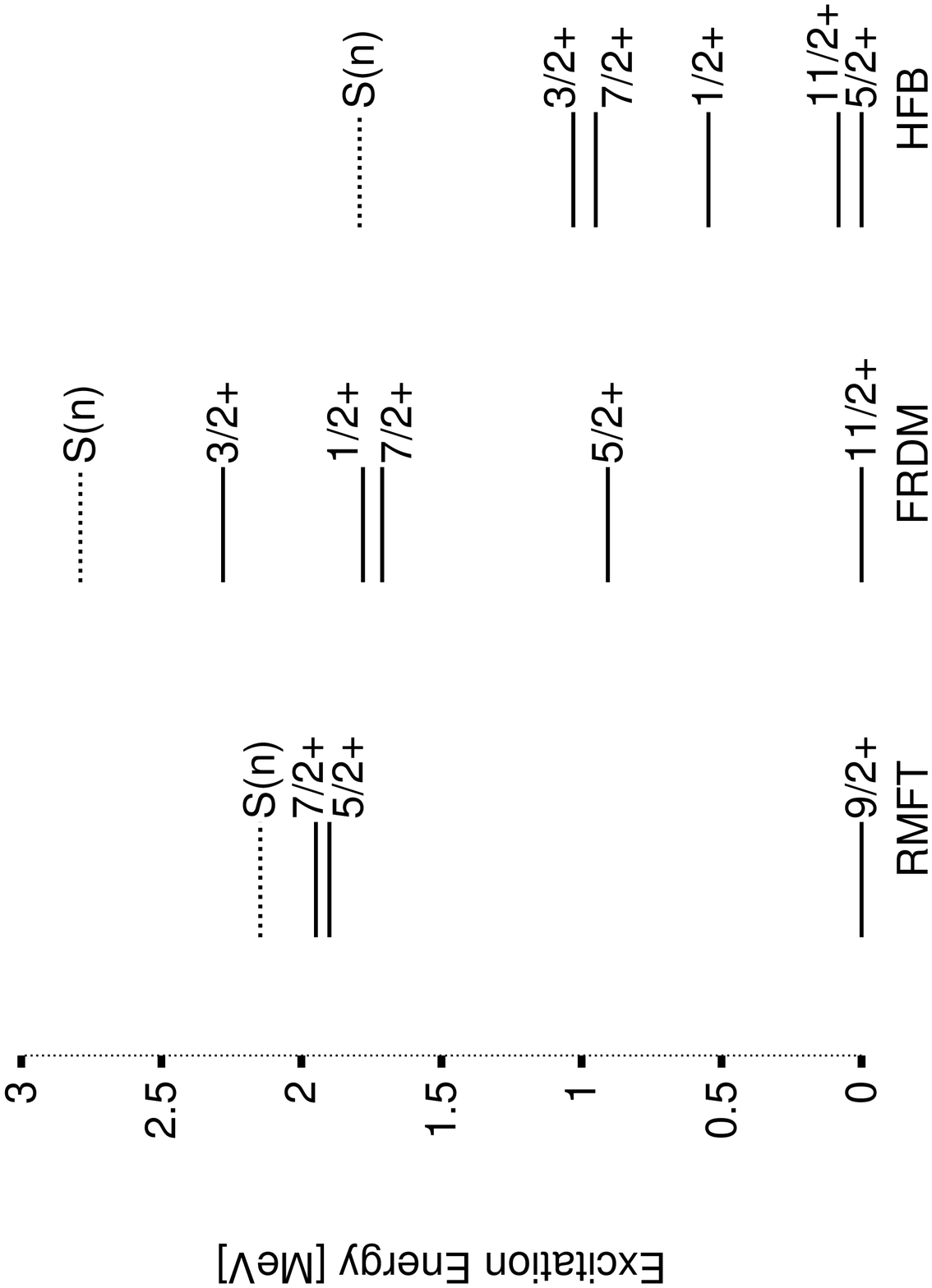,width=15cm}
\caption{\label{level2}Level schemes of $^{229}$Pb calculated
 within the RMFT,
FRDM, and HFB.}
\end{figure}

\begin{figure}
\psfig{file=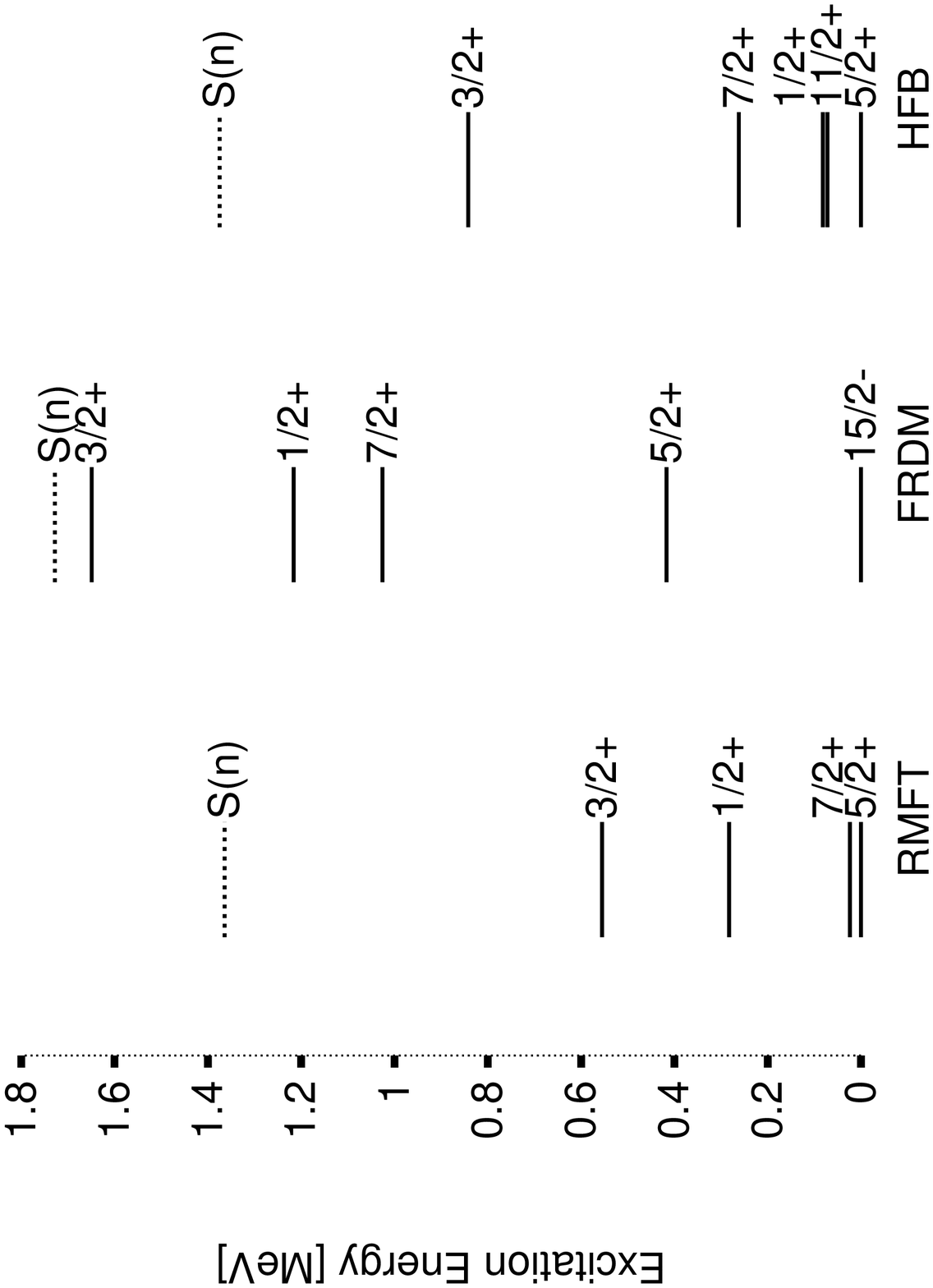,width=15cm}
\caption{\label{level3}Level schemes of $^{239}$Pb calculated
 within the RMFT,
FRDM, and HFB.}
\end{figure}

The results of our calculations are summarized in Fig.~\ref{vgl}. For
comparison, the levels from all of the models
for $^{219}$Pb, $^{229}$Pb, and $^{239}$Pb are shown in
Figs.~\ref{level1}--\ref{level3}.
The most
striking feature in Fig.~\ref{vgl} is the sudden drop over several
orders of magnitude in
the cross sections calculated with the RMFT levels in the mass range
$A=212-220$. This is due to the
lack of low spin levels which are cut off by the decreasing neutron
separation energy.  Only after the
1i$_{11/2}$ orbital
(which forms the state at lowest energy in the RMFT) has been filled
completely at
$^{222}$Pb the
cross section is increasing because low spin states become
available again. A similar gap is seen for $A=230-232$, and it is
expected that those gaps will repeatedly appear when approaching the
drip-line.
Since in some cases there are unbound low spin states
close to the threshold a small shift in the level
energies could already close such a gap. However, note that the level spacing
in the
RMFT has the tendency to increase towards neutron rich
nuclei~\cite{sharma3},
contrary to the FRDM and the HFB prediction.

The values resulting from the FRDM exhibit a smoother and
almost constant
behavior in the considered mass range.
Only a slight dip is visible for $^{220}$Pb(n,$\gamma$) since
the previously accessible 1/2$^+$ and 3/2$^+$ states have become
unbound in $^{221}$Pb. The 2g$_{9/2}$ orbital is at lower energy than the
11/2$^+$ level in this model.
Beyond $^{223}$Pb it has been filled and
at least one of the low spin states can be populated again. The known
ground state spins for the lighter isotopes are also reproduced
correctly. For higher mass numbers the cross sections are similar to the
ones obtained in the HFB model.

For mass numbers below $A=232$, the HFB capture cross sections
are always larger than those obtained in the other models.
Although the neutron separation energies are also decreasing, the $Q$-values
for the capture to the low spin states 
become even larger, because the states are moving towards lower
excitation energies.
In general, the HFB cross sections of the investigated capture reactions
exhibit a very smooth behavior with increasing neutron number.

\section{Results for Sn Isotopes}
\label{secSn}
\noindent
Proceeding in the same manner as for the Pb isotopes (Sec.\ \ref{secPb}),
we extended our investigation to the Sn nuclei.
Here, the situation is different in
two ways: Firstly, the drip-line lies at relatively much lower
neutron numbers and the r-process path is not so
far off stability, and secondly, there are more experimental data available
also for the unstable nuclei close to or in the r-process path, which makes
a test of theoretical models possible.

Again, we took the nuclear properties and density distributions from the
above described models. The strengths of the scattering potentials were
adjusted to reproduce the same value of the volume integral of 425 MeVfm$^3$
as determined from the experimental elastic scattering data on the stable
Sn isotopes~\cite{bal}.
We calculated the capture cross sections from
the stable isotope $^{124}$Sn out to the r-process path which
is predicted at a neutron separation energy of about 2 MeV~\cite{friedel}. 
As the
models make different predictions about masses and separation energies,
the r-process path is located at different mass numbers: $A\simeq 135$ for
RMFT and FRDM and $A \simeq 145$ in the case of HFB.
Contrary to the Pb isotopes for which the $p$-wave capture is the main 
contribution
allowed by the electromagnetic selection rules, the Sn cross sections
are dominated by the $s$-wave captures, due to the negative parities of the 
final states.

\begin{figure}
\psfig{file=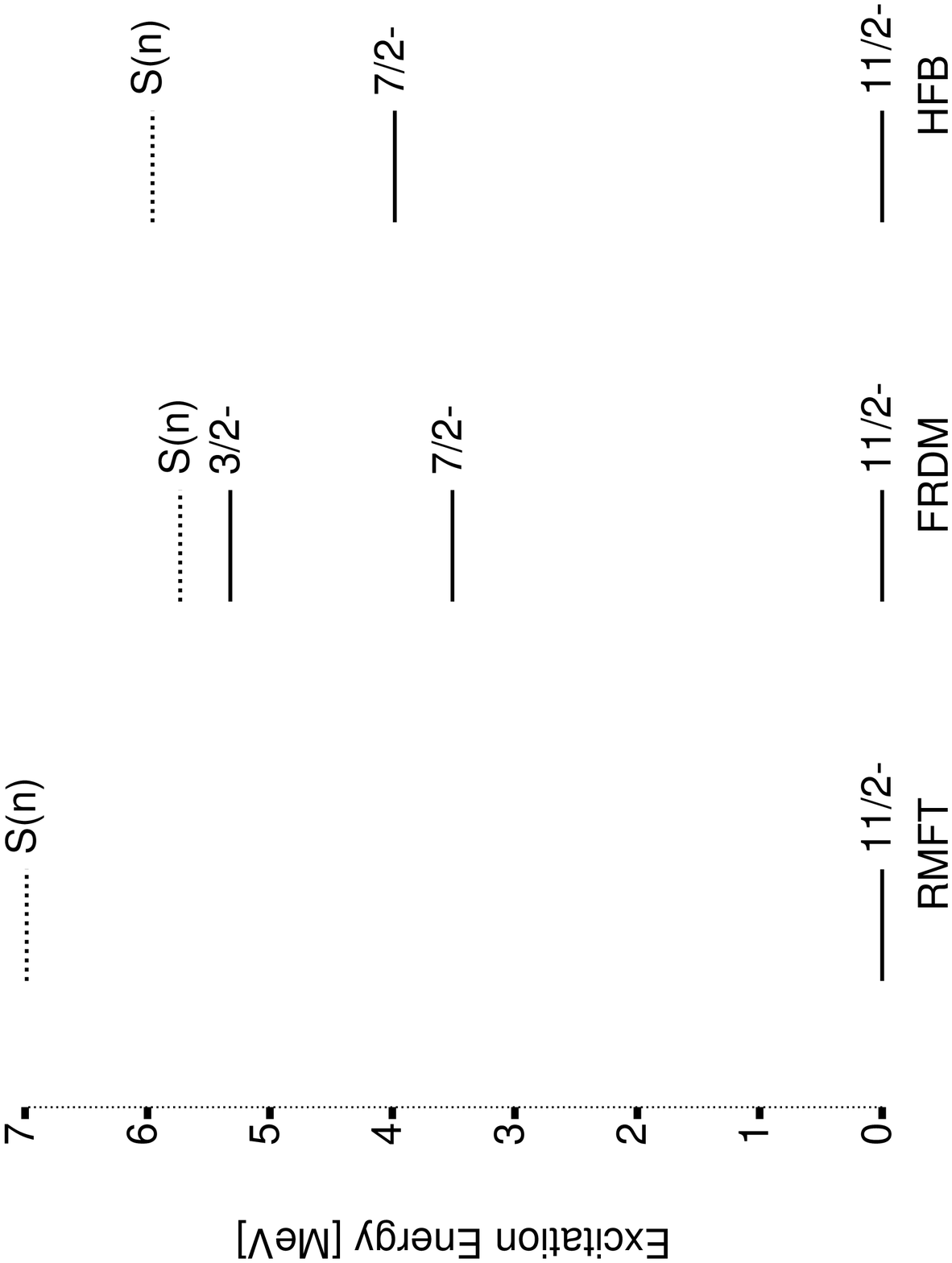,width=15cm}
\caption{\label{sn1}Level schemes of $^{125}$Sn  calculated
 within the RMFT,
FRDM, and HFB.}
\end{figure}

\begin{figure}
\psfig{file=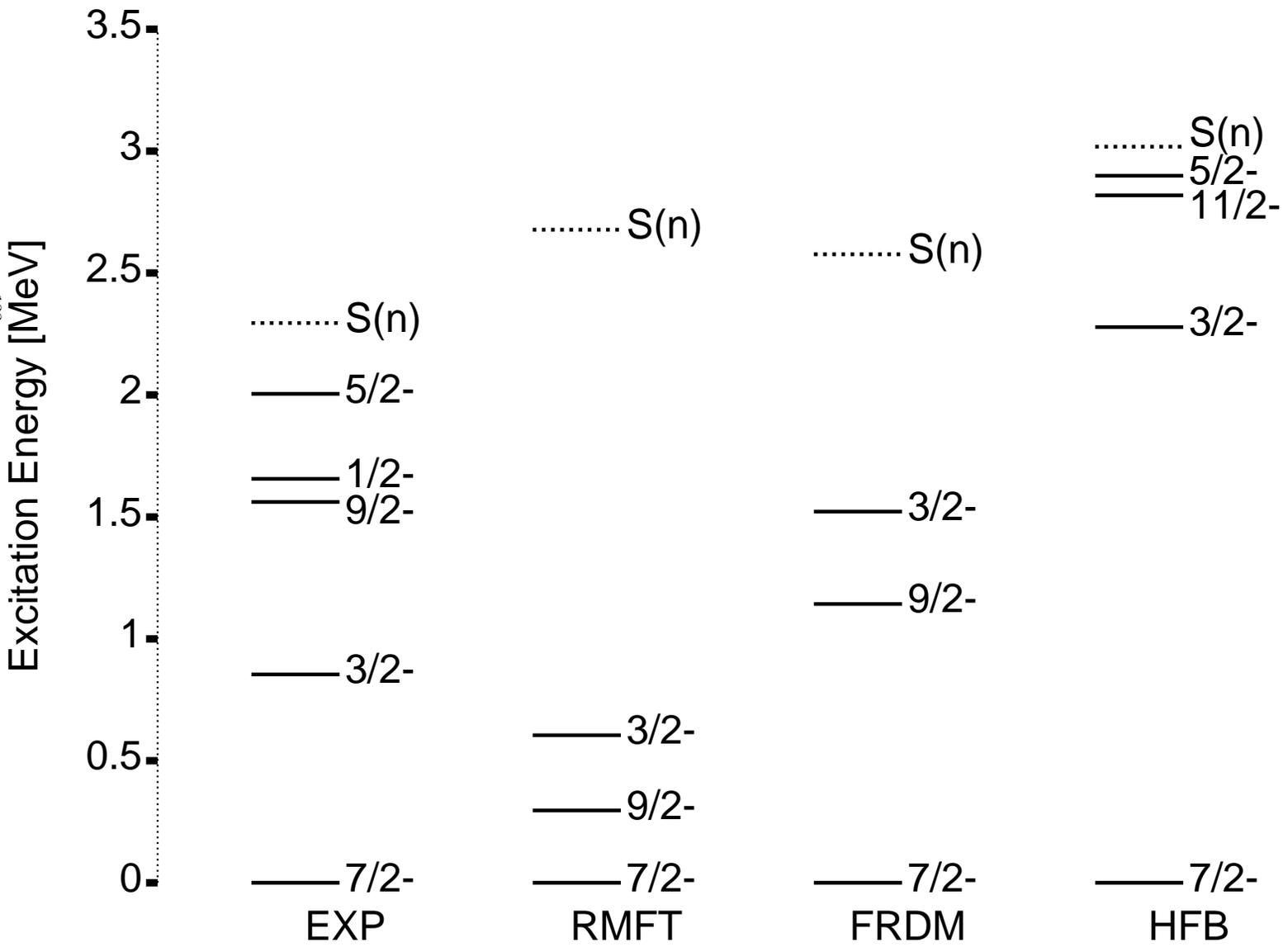,width=15cm}
\caption{\label{sn2}Level schemes of $^{133}$Sn calculated
 within the RMFT,
FRDM, and HFB. Experimental levels are taken from
Ref.~\protect\cite{hoff}.} \end{figure}

\begin{figure}
\psfig{file=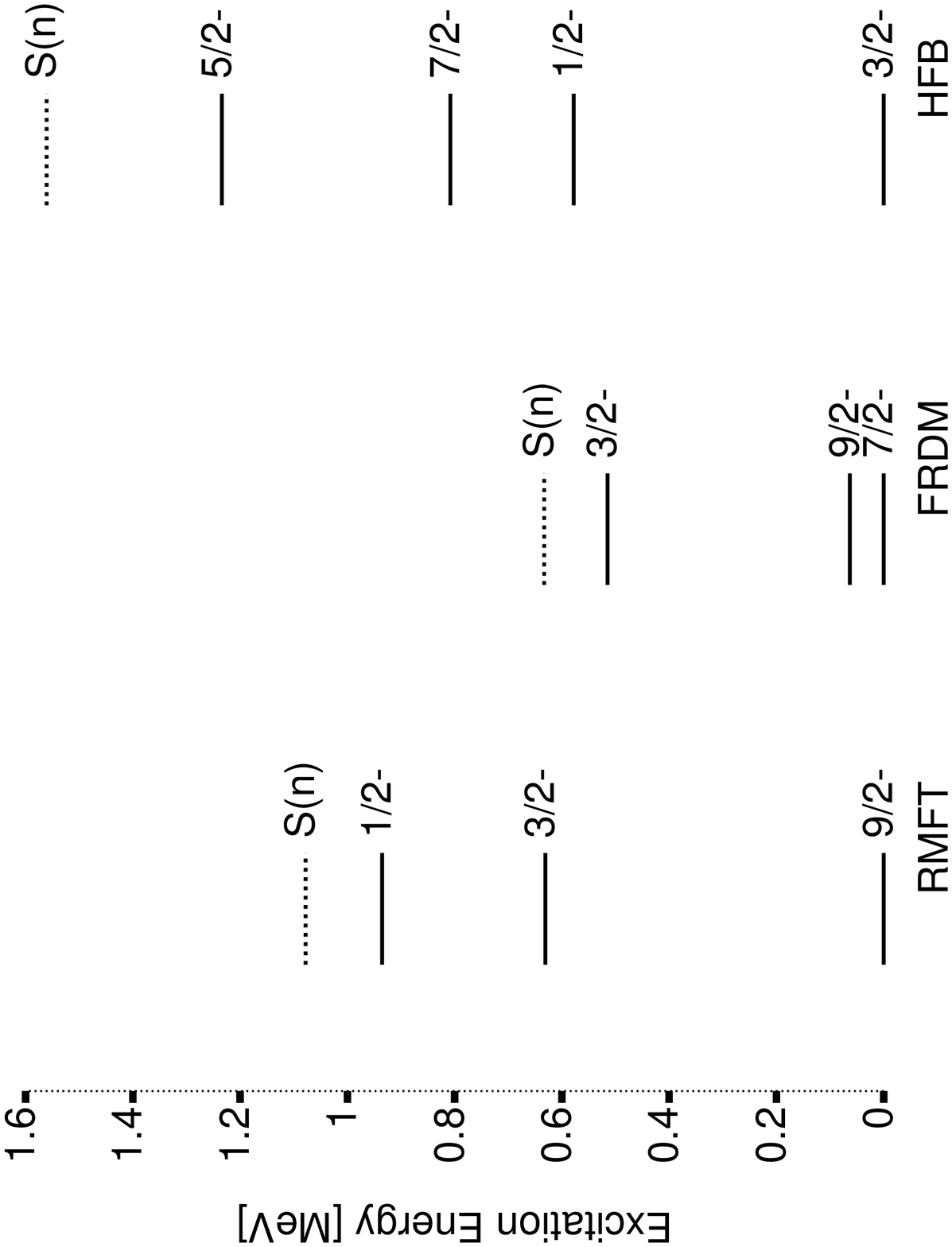,width=15cm}
\caption{\label{sn3}Level schemes of $^{141}$Sn calculated
 within the RMFT,
FRDM, and HFB.}
\end{figure}

\begin{figure}
\psfig{file=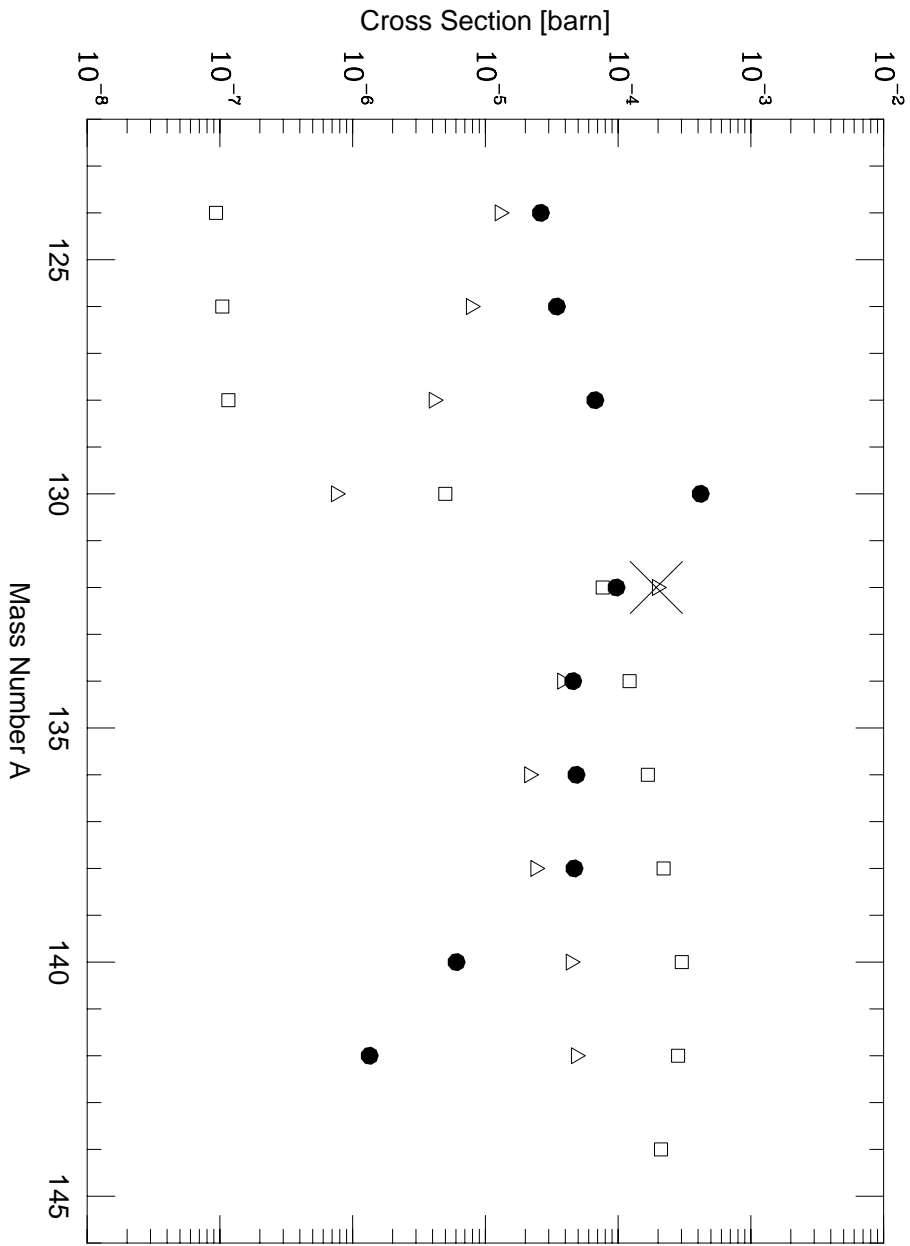,width=15cm}
\caption{\label{snfig} Same as in Fig.~\protect\ref{vgl} but for the
Sn isotopes.
The cross section resulting from a
calculation using experimental levels~\protect\cite{hoff}
is marked by a cross.}
\end{figure}

\begin{figure}
\psfig{file=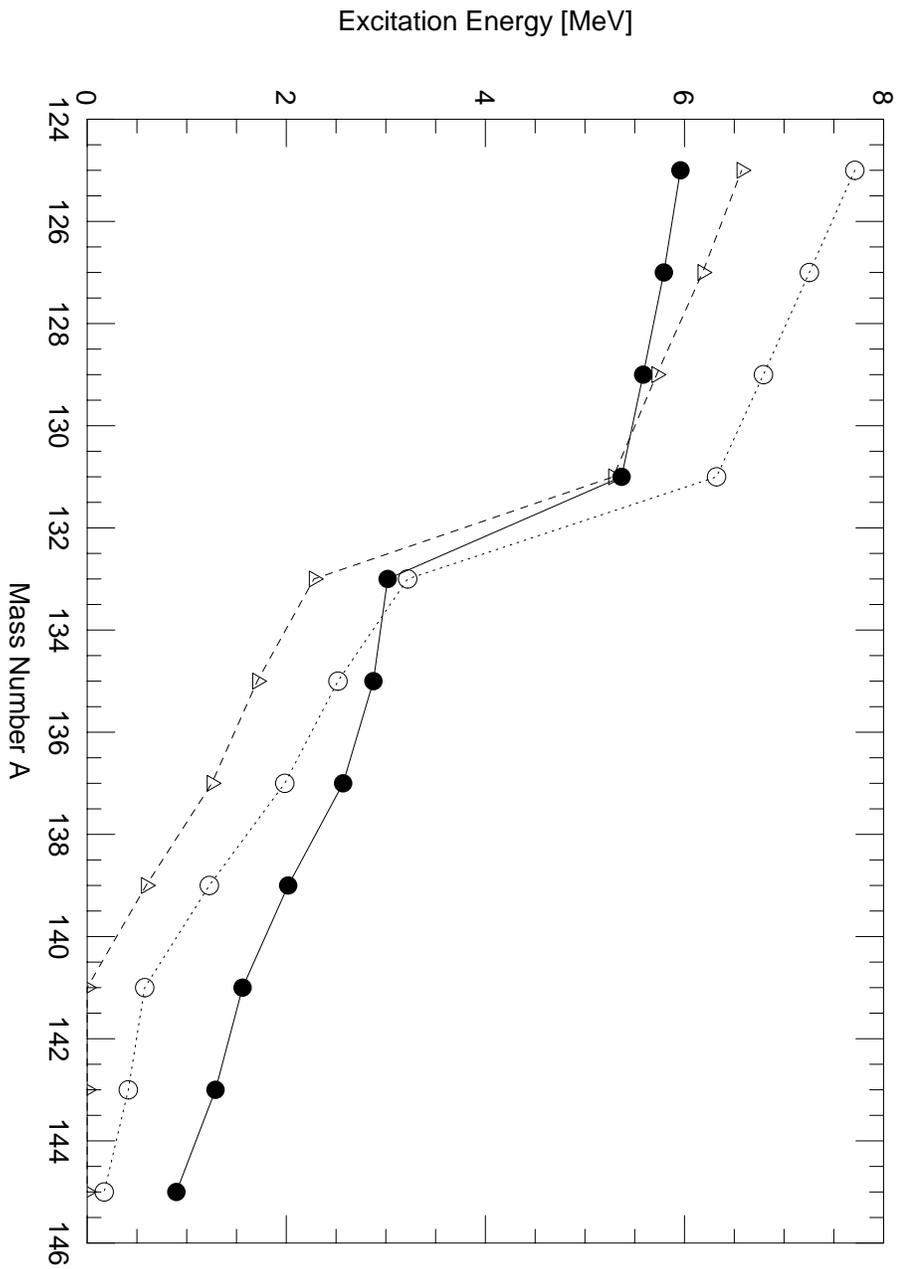,width=15cm}
\caption{\label{spinhfb}Dependence of level energies on mass number for
the even-odd Sn isotopes calculated in the HFB
model. Shown are the 1/2$^-$ state (open circles), the
3/2$^-$ state (triangles) and the calculated neutron separation energy (full
circles). The lines are drawn to guide the eye.}
\end{figure}

\begin{figure}
\psfig{file=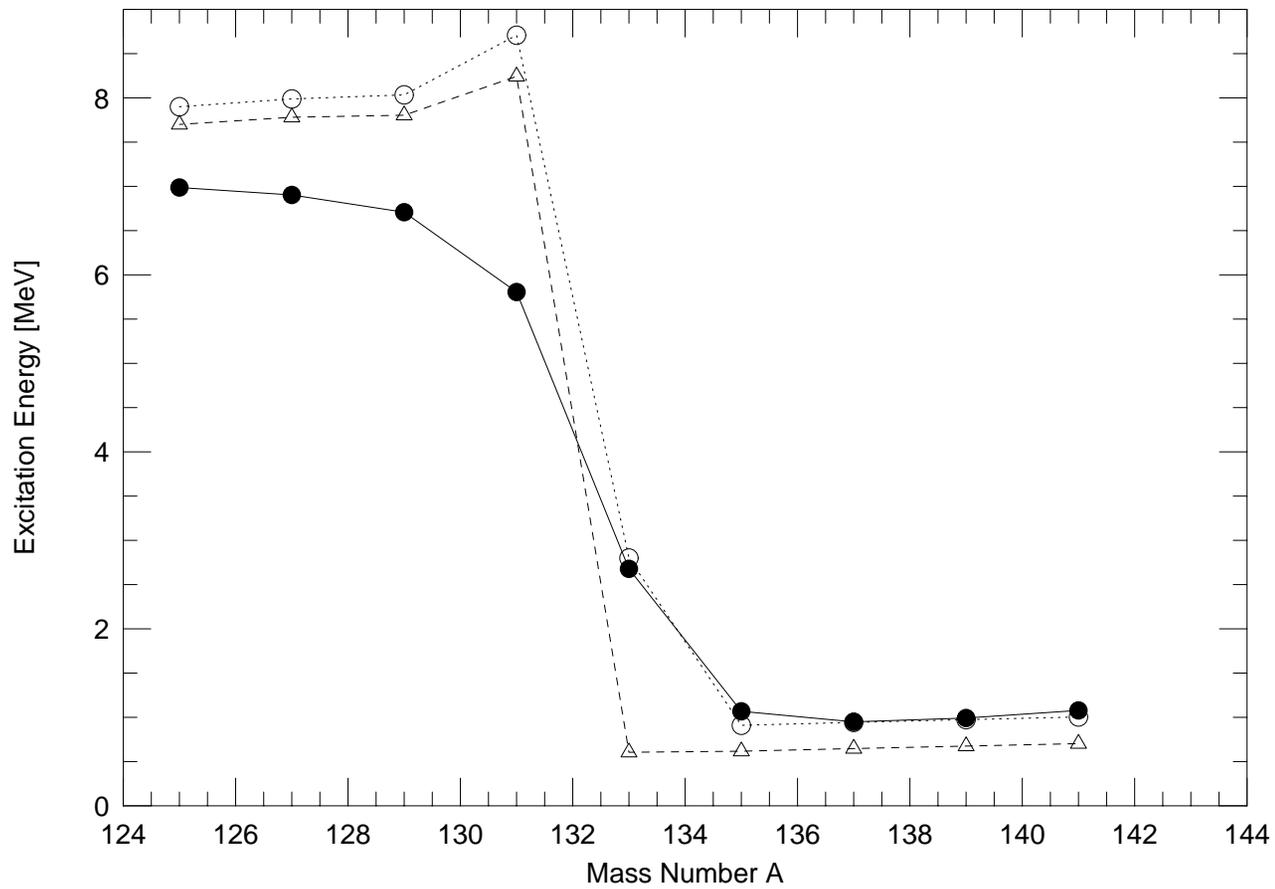,width=15cm}
\caption{\label{spinrmf}Same as in Fig.~\protect\ref{spinhfb} but
for the RMFT.}
\end{figure}

\begin{figure}
\psfig{file=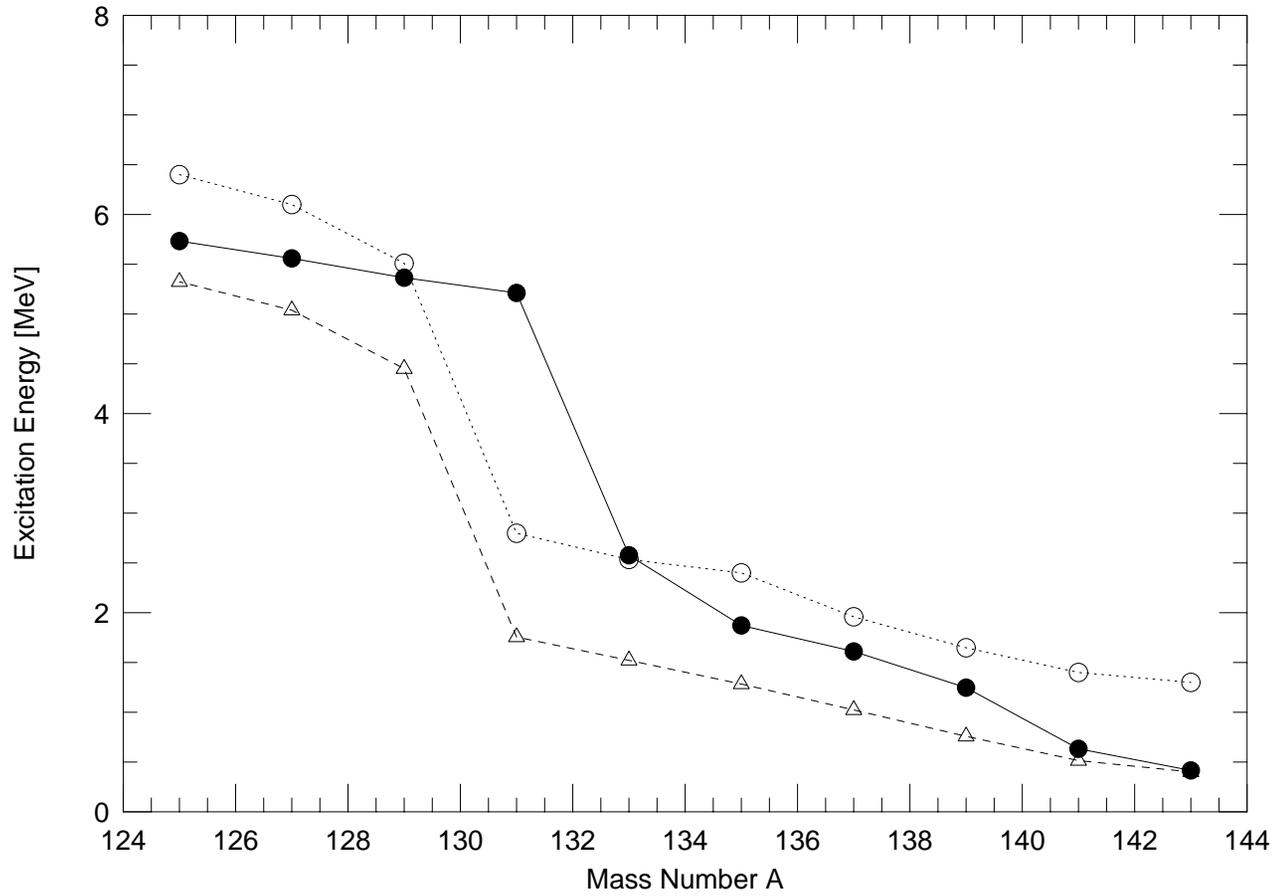,width=15cm}
\caption{\label{spinfrdm}Same as in Fig.~\protect\ref{spinhfb} but
for FRDM levels.}
\end{figure}

The level schemes of the $^{125}$Sn, $^{133}$Sn, and $^{141}$Sn
nuclei are shown in
Figs.~\ref{sn1}--\ref{sn3}, and the resulting cross sections
for all considered nuclei and models
are combined in Fig.~\ref{snfig}.
Similarly as in the Pb case,
the dependence of the cross sections on the mass number can be understood
by considering the excitation energies of the low-spin states relative to the
neutron separation energy predicted in various models
(Figs.~\ref{spinhfb}--\ref{spinfrdm}). The 3/2$^-$ state is bound in the
FRDM already at low mass number, whereas it becomes bound only at $A=131$
and $A=133$ in HFB and RMFT, respectively. Therefore, the FRDM cross sections
are larger than the ones from HFB and RMFT for $A<133$. The drop in the
FRDM cross sections beyond the $N=82$ shell is due to the fact that the
1/2$^-$ and 3/2$^-$ states slowly become unbound (see Fig.~\ref{spinfrdm}).
In the HFB model the two low-spin states move down in energy faster than
the neutron separation energy, thus providing an increasing $Q$-value and
slightly increasing cross sections (Fig.~\ref{spinhfb}).
A similar trend can be found in the
levels from RMFT, although with a less pronounced increase of the $Q$-value
(Fig.~\ref{spinrmf}).

There are no data available concerning the pure DC contribution
to the cross sections for the neutron-rich Sn isotopes. However, there is
experimental information regarding masses and level schemes. This can be
compared to theory (see Fig.~\ref{sn2}).
For the experimentally
known isotope $^{133}$Sn we calculated DC by taking the experimentally known
masses and levels~\cite{hoff} as input for the DC-calculation,
thus arriving at a pseudo-experimental value for
the cross section which can be compared to the purely theoretical predictions.
The resulting value is marked by a cross in Fig.~\ref{snfig}.
Neutron capture on $^{132}$Sn is particularly
interesting because $^{133}$Sn is predicted to be already very close to
the r-process path by the two models RMFT and FRDM.
As it turns out, however, the resulting cross sections show the closest
agreement among the investigated nuclei for this case.
All of the considered models predict the same ground state spin, a bound
3/2$^-$ state and a (barely) unbound 1/2$^-$ state (cf.,
Figs.~\ref{spinhfb}--\ref{spinfrdm}, and Fig.~\ref{sn2};
note that the mass ranges in the plots are different). However, the resulting
$Q$-value is largest in the RMFT, yielding the highest cross section. The cross
sections from the HFB and FRDM levels are smaller by about a factor of 2
because of the less strongly bound 3/2$^-$ state. The additional 5/2$^-$
state found in HFB gives only a small contribution to the total cross section
and cannot compensate for the comparatively low $Q$-value of the capture to
the 3/2$^-$ level. Nevertheless, compared to the large discrepancies
regarding other nuclei, there is good agreement in the resulting cross
sections. Therefore, this
nucleus may be a bad choice to select between the different models, but it
is reassuring in the astrophysics context that the cross sections agree so
well.

\section{Discussion}
\label{SecDis}
\noindent
In systematic r-process studies~\cite{friedel} it was found that
the r-process path is touching nuclei with neutron separation energies
around 2.5--1.7\,MeV in the Sn region and
$S_{\mathrm{n}}\simeq 1.5-0.9$\,MeV in the
Pb region~\cite{friedel}.
In our calculations for Pb (including $^{239}$Pb+n) we cover the
astrophysically relevant mass region, with the possible exception of the
HFB model. The neutron separation energies in the HFB model decrease
much slower with increasing mass number than in the other models
(cf., Fig.\ \ref{level3}), thus
not only leading to a drip-line at higher mass but also pushing the
r-process path further out. However, the most extreme path location
might still be further out by not more than two or three isotopes from
$^{240}$Pb, and therefore it is possible to extrapolate the trend seen in the
HFB calculation at lower mass numbers.
It should be kept in mind, however,
that the location of the r-process path is determined by the ratio
between neutron capture half-life and $\beta$-decay half-life.

In the following we briefly discuss the possible astrophysical
consequences of the effects
found in the cross section behavior given by the different models.
Complete r-process network calculations, which take into account all
possible reaction links and do not postulate an a-priori $\beta$-flow
equilibrium, require a large number of astrophysical and nuclear-physics
input parameters (for a detailed discussion, see e.g.~\cite{cowan}). In such
a non-equilibrium scenario, the location of the r-process path as well as
the time-scale of the r-matter flow is mainly determined by the neutron
density as astrophysical quantity, and by the nuclear-physics parameters:
the neutron separation energy $S_{\mathrm{n}}$ and the capture cross sections
$\sigma_{\mathrm{n}}$. With this, details of the r-process are depending
on the specific nuclear models used. In the following discussion
we will consider as a first estimate
only the r-process paths found in detailed studies making use of FRDM
masses~\cite{friedel} and vary the capture cross sections according to our
findings for the different microscopic inputs.

In the mass region beyond the $A\simeq195$ r-abundance peak, neutron
densities of $n_{\mathrm{n}}\simeq10^{25}-10^{27}$\,cm$^{-3}$ are required
to produce sizeable amounts of $Z\simeq80-84$, $A\simeq230-250$ r-process
isotopes very far from $\beta$-stability. After successive $\beta^-$- and
$\alpha$-decays they will form the long-lived r-chronometers $^{232}$Th and
$^{235,238}$U, and the major part of the r-abundances of $^{206-208}$Pb and
$^{209}$Bi (see, e.g.~\cite{pfeiff97}). When regarding the
$\sigma_{\mathrm{n}}$ cross sections for Pb from FRDM and HFB
(see Fig.~\ref{vgl}), very similar results are expected for the $^{230-238}$Pb
progenitor isotopes. Thus, also similar initial r-abundances for
$^{232}$Th and $^{235,238}$U will result. However, when using the RMFT
cross sections, a considerable hindrance of the nuclear flow around
$A\simeq130$ may occur which consequently would change the Th/U abundance
ratios. These neutron capture cross sections which are 5 or more orders of
magnitude smaller than the ones given by FRDM and HFB levels would
increase the life-time of a nucleus against neutron-capture by the same
order of magnitude and thus even prevent the flow to heavier elements
within the time-scales given by the astrophysical environment.

In the case of the Sn isotopes, the situation is quite different from the
Pb region. The range of astrophysically realistic $n_{\mathrm{n}}$-conditions
for producing the $A\simeq130$ r-abundances is lower, with
$n_{\mathrm{n}}\simeq10^{22}-5\times10^{24}$\,cm$^{-3}$. Hence, the r-process
path is much closer to $\beta$-stability, involving the progenitor isotopes
$^{134,136,138}$Sn only a few neutrons beyond the doubly magic nucleus
$^{132}_{50}$Sn$_{82}$. For these isotopes the Hauser-Feshbach (HF)
cross sections used so far~\cite{cowan} are of the order of 10$^{-4}$ to
$5\times10^{-5}$\,barn. According to a recent investigation~\cite{tfkl1},
the statistical model cannot be applied in that region and will overestimate
the capture cross sections. However, even if we use the experimental levels to
calculate a Breit-Wigner resonant cross section for
$^{132}$Sn(n,$\gamma$)$^{133}$Sn, we find it to be a factor of about 6 lower
than the HF cross sections.
Our present calculations would add another DC
contribution of about the same magnitude as given by
HF (see Fig.~\ref{snfig}), which has so
far not been taken into account. As a consequence of the larger total cross
section, the r-matter flow to heavier elements would be facilitated, thus
avoiding the formation of a pronounced $A\simeq134-138$ ``satellite peak'' in
the r-abundance curve sometimes observed in steady-flow calculations
(see, e.g.\ Fig.\ 2 in \cite{chen96}, or Fig.\ 5 in \cite{friedel}).
Such a signature is only indicated in the heavy-mass wing of the
$A\simeq130$ $N_{r,\odot}$-peak. It is interesting to note in this context
that the HFB model, which exhibits the weakest $N=82$ shell closure and
with this also the weakest ``bottle-neck'' for the r-matter transit in this
region (for a detailed discussion, see e.g.~\cite{klk97}), yields the
highest DC cross sections for the $A\geq134$ Sn isotopes.

Since we assumed spherical nuclei in order to be able to compare the
different microscopic models, deformation effects were not taken into
account which lead to level splitting and thus can increase the number
of accessible levels. When considering deformation our results could be
modified in two ways: Firstly, the number of bound low-spin levels could
be increased, leading to larger DC cross sections; secondly,
due to a possibly larger number of levels at and above the neutron
separation energy, the compound reaction mechanism could be further
enhanced and clearly dominate the resulting cross sections. However,
as can be seen from level density~\cite{ohurev,tfkl1} and
deformation (e.g., \cite{moell2}) studies, deformation of Pb isotopes
sets in at a mass number of about
$A\simeq 220$ and decreases already for masses beyond $A\simeq 230$.
Closer to the drip-line, the nuclei show low level densities again,
not only due to low neutron separation energies but also because of
sphericity. Lead isotopes in the r-process path (especially for components
with low $S_{\mathrm{n}}$) will therefore already
have reduced deformation and the DC -- being sensitive to the
level structure -- will give an important contribution to the total
capture cross sections. Concerning Sn, a theoretical study of the ratio of
DC over CN contributions for Sn isotopes~\cite{bal} shows that CN dominates
up to a mass number $A \simeq 130$. Moreover, deformation is predicted to set
in only at $A\simeq140$ for Sn~\cite{moellkra}.
This is supported by level density
considerations~\cite{tfkl1}, showing that the level density is too low
in this region to apply the statistical model. Therefore, depending on the
model, the r-process path lies at the border of or already well inside the region
where the DC is non-negligible and dominating.

Another source of uncertainty is the assumption of pure single-particle
states, i.e., setting the spectroscopic factors to unity. This has been
shown to be a good approximation for Pb isotopes close to stability and
it is expected to hold for neutron-rich Pb isotopes. However, a range of
0.01--1.0 for the spectroscopic factors could be realistic. This will play
only a minor role in the present comparison of different microscopic
models, as the differences in the models may be only slightly enhanced
when considering different theoretical spectroscopic factors. Nevertheless,
it will be important in quantitative calculations of abundances, invoking
complicated reaction networks.

\section{Summary}
\label{summary}
\noindent
We have shown that theoretical capture cross sections can depend
sensitively on the microscopic models utilized to determine the
necessary input parameters.
Because of low level densities, the
compound nucleus model will not be applicable in those cases.
Drops over several orders of magnitude in the cross sections -- as found
with the RMFT for Pb -- would change the position of the r-process path
and possibly influence
the formation of heavy chronometer elements,
whereas the enhanced capture rates on Sn
could have direct effects in the final r-process abundance distribution.
Deformation effects and the compound nucleus reaction mechanism
may still be of importance for the Pb isotopes and further investigations
are needed. Nevertheless, the DC will be of major importance in
the Sn region. This region is also interesting for future experimental
investigations of $S_{\mathrm{n}}$, neutron single-particle levels and
(d,p)-reactions studying spectroscopic factors.
There is also a need for improved microscopic nuclear-structure models
which can also be compared in an astrophysical context following the
successful tradition of the interplay between nuclear physics and
astrophysics.

\acknowledgements
\noindent
This work was supported in part by the Austrian Science Foundation
(project S7307--AST) and by the Polish Committee for
Scientific Research.
TR acknowledges support by an APART fellowship from
the Austrian Academy of Sciences.

%

\end{document}